\documentstyle{article}
\begin{document}

\title{The golden age of Calcutta physics: Difficulties
in reconstructing the history}
\author{Arnab Rai Choudhuri\footnote{e-mail:
arnab@physics.iisc.ernet.in}
\\
Department of Physics\\
Indian Institute of Science\\
Bangalore - 560012. India}
\date{}

\maketitle

\begin{abstract}

Classes started in the newly established Physics Department
of Calcutta University Science College in 1916. Raman, Bose and Saha
were three young members of the small physics faculty 
consisting of barely half a dozen faculty members. Within
about one decade, three extraordinary discoveries came
from these young men---Saha ionization equation in 1920,
Bose statistics in 1924, Raman effect in 1928. However,
fortunes of Calcutta University quickly got intertwined with
India's freedom struggle led by Mahatma Gandhi exactly
at the same time and the physics group got tragically 
disrupted.  Indian physics never succeeded in reaching
that height again. This paper discusses the difficulties
in reconstructing a critical history of this Calcutta
school of physics during the very short epoch of unmatched
brilliance.

\end{abstract}

\section{A look at an extraordinary epoch}

We live in an age obsessed with ranking everything:
from cinemas to washing machines, from tennis players
to universities.  Although India has one of the fastest
growing economies in the world and is striving to find
her place within the nations of the world, India's ranking
remains poor in several important spheres of human
activity. Apart from the poor showing in the Olympic
games, one other area in which India's dismal record
has become a huge embarrassment to India's intellectual
class is the ranking of universities.  No Indian university
finds a place within the world's top 100.

Just about a century ago, for about one remarkable decade, an
Indian university reached a height which no Indian
university afterwards succeeded in reaching. It was
Calcutta University under the leadership of the
visionary Vice-Chancellor Sir Asutosh Mookerjee. That
was a time when the world ranking of universities had not
yet begun and we do not know what the rank of Calcutta
University at that time would have been.  However, judged
by any reasonable yardstick, Calcutta University surely
would have ranked among the world's greatest for a brief
period of unmatched brilliance. Although Mookerjee
managed to attract outstanding faculty in as diverse
fields as philosophy, history and art, here we are 
concerned with the Physics Department of the University
during the golden period. Mookerjee was a brilliant
mathematician himself and, at a very young age, published a string
of original papers in mathematics in leading journals, 
attracting international attention.  As there was very little 
scope of a career in mathematics in India of that time,
Mookerjee had to take up the legal profession before
being appointed to head Calcutta University.  Although
Mookerjee could not pursue mathematics research in his later life,
he maintained a lifelong interest in mathematics and
physics. When Mookerjee took over the reins of Calcutta
University, universities in India were 
primarily bodies for conducting examinations.  Mookerjee
wanted to create post-graduate departments where faculty
members would carry on research.  India was under the
British rule and Mookerjee knew that financial support
from the Government would not be forthcoming.  He
established the University Science College with
donations from rich Indians, which enabled him to
establish a few professorships.

An account of how Mookerjee built up the Physics Department
of Calcutta University Science College
almost reads like a fairy tale.  He wanted to attract faculty
who would teach modern topics in physics and simultaneously carry on
research.  There was no tradition of physics research
in India at that time, apart from solitary workers like
J.~C.\  Bose at Presidency College (who was no longer working on physics at
that time after his fundamental studies of radio waves). 
Mookerjee started looking for young
people who had the potential for growing into outstanding
physicists. He knew of a 26-year-old officer in the Finance
Department, who was passionate about physics and had already
published about a dozen papers in top international journals
by carrying on research in his spare time.  Mookerjee wanted to 
get him for the most prestigious chair of the fledgling Physics
Department---the Palit Professorship.  However, Mookerjee could offer
him only Rs.\ 600 against his salary of Rs.\ 1100.
Would he be willing to take up this professorship?  The young
man, C.~V.\ Raman, jumped at the offer. Mookerjee also needed
younger persons to man the department.  As it happened, the
batch which completed master's degree in the year 1915 was an
exceptional batch.  Mookerjee called three bright boys of
that batch for a discussion$^1$. Only one of them, Sailen Ghose,
was a student of physics. Although the other two, Satyendra
Nath Bose and Meghnad Saha, were students of mixed mathematics
(what we would now call applied mathematics), Mookerjee knew
that they were interested in physics.  Mookerjee asked the
three boys if they could teach the modern topics of physics
which had never been taught in any Indian university. Saha
was assigned to teach quantum theory and Bose was assigned
to teach relativity.  Sailen Ghose, who was a good experimenter,
was given the job of designing the laboratory course and setting
up the experiments.  

Before classes started 1916, there was trouble. Although
Sailen Ghose managed to acquire laboratory equipments and
set up the laboratory, he could never formally join the
department. He had connections with revolutionary groups
fighting the British imperialism. Police found clues about
this and raided his home when he was away. It appeared that
he would be sent to the British penal colony of the Andaman
Islands if caught.  Ghose fled India in a ship bound for
Philadelphia disguised as a Muslim crew member, thus putting
down the curtain on what appeared to be a very promising
career in physics. Ghose, however, later became a highly
respected figure in the arena of international political
activism.  The short autobiography he wrote for the American
magazine {\em Asia} in 1927 gives a fascinating account
of the Calcutta academic world in his youth$^2$.

Bose and Saha were among the first to join the new Physics
Department in 1916---at the monthly salary of Rs.\ 150. Raman joined
about a year later when he had made up his mind to quit the
Finance Department.  The person to whom the other named
chair (Rashbehari Ghosh Professorship) was 
offered---Debendra Mohan Bose---was in Germany when the
First World War broke out. He was forced to stay there till the end of the
War and could join Calcutta University only after the War
ended.  
Raman, Bose and Saha were members of the small new 
Physics Department with barely half a dozen faculty members.  
Within about a decade, three spectacular physics discoveries
emerged out of this group---Saha ionization equation in
1920, Bose statistics in 1924, Raman effect in 1930.
Although Bose had shifted to Dacca University a little
bit before his work on Bose statistics and, in that sense,
that work should technically not be counted as a work from
Calcutta University, Bose always had strong links with 
Calcutta University and eventually returned there again
in later life.  There were very few physics groups around
the world---perhaps no group outside England and Germany---which
could boast of three physics discoveries of this class in that
decade.  In many of the present-day surveys, Harvard University
appears at the top.  There was not even one physics discovery
of this class made at Harvard University during the same period. 
Although the achievements of Raman, Bose and Saha dwarfed
the achievements of their colleagues, it should be emphasized
that some of their colleagues such as Sisir Kumar Mitra and
Debendra Mohan Bose also made very important contributions
to physics during the same period. 

It seemed that this extraordinary decade heralded a new dawn
for Indian physics.  But that was not to be.  That brilliant
dawn quickly faded away and a fourth physics discovery of that class
has not come out of India in the next 85 years!  There were
some obvious external reasons why academic activities of
Calcutta University got derailed in a big way. Mahatma Gandhi
returned to India in 1915 and the infamous Jallianwala Bagh
massacre took place on 13 April 1919---exactly around the time
when the fledgling Physics Department of Calcutta University
was trying to find the ground under its feet.  The golden
age of Calcutta physics almost exactly coincided with the
golden age of Indian freedom movement when the Muslim question
still had not raised its ugly head and Indians were fighting
together for their country.  In fact, this coincidence might
not be accidental.  Raman, Bose and Saha have all categorically
stated on various occasions that patriotic fervour was one of
the driving forces behind their science.  They had to prove
to the world that Indians could compete with white men in
science.  Saha had difficulties in getting a government
job in his youth because he was perceived as a sympathizer
of the revolutionary groups fighting the British Empire.
Raman was found weeping when he received the Nobel Prize
and gave a moving account of it himself:
\begin{quote}
When the Nobel award was announced I saw it as a personal
triumph \ldots But when I sat in that crowded hall and I saw
the sea of western faces surrounding me, and I, the only
Indian, in my turban and closed coat, it dawned on me that
I was really representing my people and my country \ldots
Then I turned around and saw the British Union Jack under
which I had been sitting and it was then that I realized
that my poor country, India, did not even have a flag of
her own---and it was this that triggered off my complete
breakdown.$^3$
\end{quote}
Although Mookerjee always expressed the opinion
that students can serve their country best by excelling in
their work and was against students boycotting
classes, it was an open secret that he and many professors
of the University were sympathetic to the national cause. The
British administration decided to punish the University for
its insubordination.

All government funds to Calcutta University were drastically
cut.  There was a period when the University was unable pay
the salary of its faculty for several months. At the height
of the crisis, in a Senate meeting on 8 December 1922 to discuss
a financial offer from the government with humiliating conditions,
Mookerjee spoke stirringly: ``If you give me slavery in one 
hand and money in the other, I despise the offer \ldots
Our post-graduate teachers would starve themselves, rather
than give up their freedom \ldots I call upon you, as members
of the Senate to stand up for the rights of your University \ldots
Freedom first, freedom second, freedom always: nothing else
will satisfy me.'' In a letter dated 26 March 1923, Mookerjee
turned down the offer of another term of Vice-Chancellorship
with humiliating conditions.  Then he suddenly died on 25 May
1924 at the age of 59. In the meeting at Calcutta University
to condole his death, Lord Lytton, Viceroy of India who was
also the Chancellor of Calcutta University, said:
``Let each of us severally resolve that this cherished creation
of his life shall not suffer because he has left us \ldots
let all differences be forgotten, all mistakes forgiven, let
us resolve to build over his ashes a temple of reconciliation.''$^4$
However, it was already too late.  Although Mookerjee hoped
that the ``post-graduate teachers would starve themselves'',
many of them had already left Calcutta for greener pastures
by the time of his death.
Bose left for Dacca University in 1921 and Saha for Allahabad
University in 1923. Calcutta University could never regain
its lost glory.

Saha was abroad at the time when the financial crisis had
first struck Calcutta University. Saha must have been aware
of this crisis and applied for a grant to the British High Commissioner
of India for his research on his return to India.
The mood of the period can be gauged from the letter dated
9 February 1921 that Mookerjee wrote to Saha at that time: 
\begin{quote}
I wish you had applied to your Alma Mater and not to the
High Commissioner.  We are in great financial crisis here
on account of Non-Cooperation movement, but you may rest
assured that so long as it is practicable, your Alma Mater
will not be slow to help you \ldots I trust you will not
hesitate to serve your Alma Mater when you return.  I was
deeply grieved to hear that Dr.\ Jnanendra Ghosh had decided
to give up his Alma Mater and accept service in Dacca University.
When will the children of our Alma Mater realise that it is
absolutely necessary for all of them to stand by Her at the
most critical period of the history of Her development?$^5$
\end{quote}
While reading this letter, we can almost hear the anguished
voice of that proud visionary who built Asia's most
outstanding university of that time and then saw the creation
of his dreams crumble before his own eyes. Physics research
in India could never reach the height that it reached during
the extremely short golden period of Calcutta University!

Whenever I have discussed this extraordinary history with my 
non-Indian friends, many of them wanted to know if there
exists a coherent account of this history that I could recommend
to them. I did not know of any book that covers the history
of this extraordinary epoch of physics research in Calcutta
comprehensively.  There are authoritative
biographies of Raman, Bose or Saha, some of which I
shall list later in \S~3.  However, most of these
biographies have been written
specifically for Indian readers.  For example, a scholarly 
biography of Bose begins by referring to Bengal, Calcutta and Dacca
in the opening paragraph.  An Indian reader would be expected to
know that Bengal was a province of the British Indian Empire
comprising roughly of present-day Bangladesh and the state
of West Bengal in India, and that Calcutta and Dacca were the
first and second largest cities of that province. A non-Indian
reader who does not have this background of geography and
history would be quite a bit lost. Also, one can have a full 
picture of that extraordinary epoch only from a study focussing
on that epoch rather than from individual biographies,
setting that epoch as a landmark in modern India's search
for self-expression against the backdrop of the 
freedom movement. To
the best of my knowledge, no such study focussing on this
epoch that would be easily accessible to non-Indian readers
exists.    

Apart from the dramatic way in which the Physics Department
of Calcutta University rose and fell, there are other elements
of high drama connected with our history. I give two examples.
The first example concerns Raman, who hailed from the Tamil
country in the deep south of India and had come to Calcutta in
1907 with a job in the Finance Department. While on his way back 
from work by tram one day, he noticed the signboard   
of Indian Association for the Cultivation of Science, which
was started by Mahendra Lal Sircar, a successful
medical doctor, about a quarter century
earlier for the `cultivation of science', but remained a dormant
place in which, to Sircar's great disappointment, nobody took
any interest.  Raman discovered this sleepy place stocked
with various kinds of scientific instruments where nobody
was working, and Amrita Lal
Sircar, the son of Mahendra Lal Sircar who had passed away
by then, was believed to have risen 
to embrace Raman with the words ``All these years
we have been waiting for a person like you'',  and made all the
facilities available to him.  It was the Indian Association
for the Cultivation of Science where Raman carried out his
initial research at weekends and in the evenings that attracted
the attention of Asutosh Mookerjee and brought the offer of
Palit Professorship. The second example is about the
way Bose's famous work got published.  When Bose had difficulty
publishing his epoch-making paper in {\em Philosophical Magazine}
where he had already published three papers, he sent the paper
to Einstein with a covering letter.  Einstein immediately
understood the revolutionary nature of the paper, himself translated
it into German from English and arranged for its publication
in {\em Zeitschrift f\"ur Physik}. Later on, Einstein extended
Bose's analysis for photons to particles with mass.

Certainly Raman, Bose and Saha belong to the whole world 
and not just to India.  Their story ought be told in a way
that is accessible to readers outside India.
Even though the history of the physics community in Calcutta at the
time of the establishment of the Science College is so rich
in human drama, the important question before us is whether we
can reconstruct that history reliably and critically with
the help of materials available to us at the present time.
We look at this question in this paper.

\section{The role of our epoch in the growth of modern science
in India}

Raman, Bose and Saha occupy a peculiar place in the history of
the growth of modern science in India. They were certainly not
the first Indian scientists to receive international attention.
That honour should go to the physicist J.~C.\ Bose and the chemist
P.~C.\ Ray---both teachers at Presidency College in Calcutta---who 
rose to scientific eminence towards the
end of the nineteenth century.
Then, just around the time when the Physics Department
of Calcutta University was being planned, Srinivasa Ramanujan
dazzled the world of mathematics like a brilliant short-lived
comet. All these scientists depended on the West in very crucial
ways for their creativity.  (J.~C.) Bose and Ray had their
initial training in research in London and Edinburgh respectively.
Although Ramanujan wrote his first paper before leaving India,
his genius blossomed only when G.~H.\ Hardy got him to Cambridge. Compared
to them, Raman, (S.~N.) Bose and Saha were completely self-made.
None of them ever had a proper research `supervisor'.  All of them
figured out on their own what they wanted to do.
The famous discoveries of Saha and Bose were made before they ever
stepped out of India.  While Raman had been abroad after his
early research made his reputation, that international exposure
was probably not too crucial in shaping his subsequent research
path leading to the discovery of the Raman effect.  

Although the Western influence was so important in moulding
the careers of (J.~C.) Bose, Ray and Ramanujan, they were often
perceived by their contemporaries as the culmination of the 
Indian tradition.  Both (J.~C.) Bose and Ray were generally
referred to as {\em Acharyas}---an epithet for great teachers in the
Indian cultural tradition.   In the case of J.~C.\ Bose, after
his early brilliant work on radio waves, when he shifted to studying
responses in plants, it became easy to connect him to the Indian tradition
which perceived a unity in nature. Although Ray's chemistry research
could not be connected to the Indian tradition that way, Ray also
carried out pioneering research on the history of Hindu chemistry.
In Ramanujan's case, the lack of importance he often attached to 
mathematical proofs could easily be ascribed to his incomplete training in
mathematics in his formative years$^6$. However, since mathematical
proof was historically given less importance in the Indian mathematical tradition
compared to the Greek tradition, one could connect the peculiarities
of Ramanujan's genius to his Indian heritage. Ironically, although
Raman, (S.~N.) Bose and Saha were the first modern Indian scientists whose
careers were not shaped by their interactions with the West as in the case
of their predecessors and, in that sense, they were more indigenously
Indian, they were the first generation of Indian 
scientists who were generally not viewed as being connected with
the Indian tradition.  They were simply regarded as international
scientists belonging to the global tradition of science.  

After these comments on how Raman, Bose and Saha stood in relation
to their predecessors, we come to the more important question:
how do they stand in relation to their successors?  The short
brilliant burst of scientific creativity with which they had
been associated did not lead to the establishment of a school
of physics research. It is vitally important for us to address
the question why this extraordinary phase of scientific
creativity disappeared as suddenly as it appeared.  Was it
that conditions in India at that time were such that it was not 
possible for such a creative phase to continue for too long?  Or do
we have to `blame' Raman, Bose and Saha in some way that, in spite 
of their extraordinary achievements, they did not provide the
right kind of leadership for the growth of Indian physics?  
Since Raman, Bose and Saha have been icons of science in India,
it has been almost a taboo to raise such a question.  Nearly
all the Indian authors who wrote on them refrained from analyzing
this issue. Now that we can view them from a historical distance,
perhaps it is time to analyze this vital issue objectively. 

Amal Kumar Raychaudhuri, the most outstanding physicist to
come out of Calcutta in the next generation, is remembered for
the Raychaudhuri equation which played a crucial role in proving the
singularity theorems in general relatively.  Raychaudhuri, who had seen Bose
and Saha closely, wrote:
\begin{quote}
The present writer passed out of the University in 1944. I remember how those who
were a few years senior discouraged me about taking up research. The feeling was that
one should take up any job whatsoever that he may be able to get rather than enter into
a field where prospects were simply dismal. To make matters worse an idea went around
that scientific research is the business of supermen like Bose and Saha and an ordinary
Indian should not be foolish enough to aspire after doing anything worthwhile. Strangely
neither Bose nor Saha did anything to counter such absurd ideas---I wonder whether they
actually relished it!$^7$
\end{quote} 
Note that Raychaudhuri is referring to the year 1944---only
slightly more than a decade after the glorious era of Calcutta
physics when many of the heroes of the glorious era were still
very much around.  How could the mood change so much in such a
short time? We have to keep in mind that the national mood as
a whole was much gloomier and darker at that time compared
to the forward-looking 1920s.  That was the time of the
Second World War.  Bengal was recovering from the terrible
man-made famine of 1943.  Communal riots between Hindus
and Muslims started becoming a frequent occurrence. Still
we get a feeling that this societal mood does not fully
explain Raychaudhuri's pessimistic outlook. Even at the
darkest of times, human spirit has an urge to conquer the
circumstances and rise above them. 
Could it be that Raychaudhuri was a natural pessimist and
his view did not really reflect the view of his generation?
Raychaudhuri himself and many of his contemporaries had been
my teachers in Presidency College in the 1970s.  
With several of them, I had 
detailed conversations about their student days.  I can assert with 
confidence that the quotation from Raychaudhuri quite accurately
reflects the  mood of his generation.

If there is a sudden burst of scientific creativity in a country
which did not have a tradition of scientific research, there
are opposing examples of that leading to the establishment
of a school of research and also that {\em not} leading to the establishment
of a school of research.  Let me give two such contrasting examples.
In the second half of the nineteenth century when America was
in the backwaters of scientific research, Josiah Willard Gibbs,
a solitary genius working at Yale University, made very deep and
profound contributions to theoretical physics.  He was an
intellectual recluse who did not leave followers behind him
and his influence did not give rise to any school of physics research
in the USA.  Contrast with this the case of the Russian physicist
Lev Landau, a few years younger than Raman, Bose and Saha.
Although there had been some earlier Russian physicists who
made important contributions to theoretical physics, it was
the charismatic personality of Landau, through his teaching and
research mentorship, that inspired the whole next generation of
Russian physicists and established a very strong school in
theoretical physics.  If we want to compare the Indian situation
with the Russian situation, we have to keep one important factor
in mind. It was not easy for a Russian physicist to travel
abroad.  So the brightest people stayed in the country and 
produced their best science there.  On the other hand, brain
drain had a serious impact on the growth of Indian science as
countries like the USA became more welcoming to scientific
emigres. This, however, does not explain the mood of despondency
expressed in the quotation from Raychaudhuri. 
   
When would we expect that a sudden burst of scientific
creativity would lead to the establishment of a vibrant scientific
tradition?  I would like to humbly propose the following
answer. Only when there is a community of many active competent
scientists, we expect that a few of them may make spectacular
discoveries, giving rise to a strong scientific tradition.
In a given population, there may be a small handful of unusual
geniuses who would rise to scientific eminence even under the
most adverse circumstances.  Presumably Raman, Bose and Saha,
as well as Raychaudhuri in the next generation,
belonged to this class. But one cannot have a school of science
only with such unusual individuals. There would be a much larger
number of persons in the population who have the intrinsic
ability to become competent scientists, but who blossom into
competent scientists only if they receive appropriate guidance
and encouragement in their formative years. A community can have
a strong school of scientific research only when this second
category of persons are able to realize their full potential.
When we consider the generation after the generation of
Raman, Bose and Saha, we clearly perceive it as a generation
of missing physicists.  I can only talk about that generation
in Calcutta, because I am not sufficiently familiar with that
generation elsewhere in India.  Those of us who had the privilege
of being taught physics by Shyamal Sengupta, Rashbehari Chakrabarti 
and Hemendra Nath Mukherji in Presidency College in the 1960s and
1970s would unanimously agree that they had the potential to become 
outstanding physicists if circumstances were different in 
their youth. Their command over physics would easily surpass that
of many professors in India's top physics departments today.
In spite of the spectacular achievements of Raman, Bose and
Saha, why did Calcutta not have an intellectually stimulating
atmosphere for physics in the following years?  Why many would-be
physicists never really blossomed into successful researchers? What caused this
generation of missing physicists?

We now come back to the question whether we have to `blame' Raman,
Bose and Saha for this.  Since we are mainly talking about 
Calcutta now and Raman left Calcutta in 1933 to take up the
Directorship of Indian Institute of Science in Bangalore, let
us leave Raman out of our reckoning right now.  Although Bose
and Saha spent several years in Dacca and Allahabad respectively,
both of them eventually came back to Calcutta University---Saha
in 1937 and Bose in 1945. There was a period of about seven
years (1945--1952) when Saha and Bose were both in the physics
faculty of Calcutta University. Many of our outstanding teachers
in Presidency College went through Calcutta University exactly
during this period.  As I have already pointed out, there have been
supremely great physicists like Gibbs who simply did not have the 
type of personality to inspire the next generation that Landau
could do.  If Bose and Saha also like Gibbs did not have the right type 
of personality, then we can feel sorry that physics did not take
root in Calcutta in a way we would have liked, but we can hardly
blame Bose or Saha for not being Landau!
We should blame them only if they did something that was
detrimental to the growth of physics in Calcutta.  The quotation
from Raychaudhuri obliquely hints that Bose and Saha indeed have to be
blamed, to some extent at least, for 
what was happening (or rather what was not happening)
in the physics world of Calcutta.

There is now a genre of biography-writing which is essentially
slinging mud at great persons.  The biographer digs out all
kinds of unknown negative secrets in the life of a great man
to cut him down to size.  I personally have never been an admirer
of this genre of biography-writing. But uncritical iconography
is not the viable alternative.  Raman, Bose and Saha were scientific
giants of such colossal stature that even if our analysis shows
that they have to be blamed for their negative impact on the
growth of Indian physics, that will not diminish the brilliance
of their extraordinary achievements.  However, an objective analysis 
of this should have much broader implications for the whole
subject of history of science.  Such an analysis would enable
us to understand better the circumstances under which a tradition
of scientific research may develop and flourish, as well as what
may stifle such a tradition. The important question is whether
we have sufficient source materials available to us at the present
time to reconstruct a history of the glorious epoch 
of Calcutta physics in sufficient
detail to make a proper objective analysis possible.   We now
take stock of the available source materials and address this
question.

\section{In search of source materials}

Since less than a century had elapsed after the epoch in which
we are interested and many persons who had seen Raman,
Bose and Saha in flesh and blood are still alive, one
may think that reconstructing a history of that epoch should
not be too difficult.  But as I started looking for source
materials, I quickly realized that this is a much more
challenging job than what I initially expected.  
The scientific papers of all three are easily accessible, since
their important papers appeared in standard journals and
the collected papers of all of them have also been published$^{8-10}$.
Saha had written a considerable amount apart from his
scientific papers.  These writings also have been collected$^{11}$.
The collected Bengali writings of Bose and Saha have also
been published$^{12-13}$. 

We do know the broad outlines of the major events in
the lives of our three protagonists---Raman, Bose
and Saha.  Soon after their deaths, short sketches of
their lives appeared in {\em Biographical Memoirs of
the Fellows of the Royal Society} written by persons who
knew them well and knew about their science (D.~S.\ Kothari
wrote on Saha$^{14}$, S.\ Bhagavantam on Raman$^{15}$, J.\ Mehra on Bose$^{16}$). 
The primary sources of information about the lives of Bose and
Saha are the short biographies penned by Santimay Chatterjee,
a Calcutta-based physicist, and his writer-wife Enakshi
Chatterjee$^{17-18}$.  Santimay Chatterjee had done research under
Saha's supervision and had known both Bose and Saha
personally.  Although the biographies were written after
the deaths Saha and Bose, Chatterjees got quite a lot of
their information from the family members of Saha and Bose.
Afterwards Santimay Chatterjee, in collaboration with others,
prepared a fuller biography of Bose on the occasion of
his birth centenary$^{19}$.
The most detailed account of Raman's personal life can be
found in the biography by Uma Parameswaran$^{20}$, who was
the granddaughter of Raman's elder brother.  She got
much of her information from family members---especially 
from Raman's wife Lokasundari, who lived
for several years after Raman's death and whom Parameswaran
knew intimately. S.\ Ramaseshan, Raman's nephew who carried on
research under Raman's supervision, also wrote several
articles on Raman as scientist and man$^{21}$. 
A monumental scientific biography of Raman was
written by G.\ Venkataraman$^{22}$, who also wrote excellent short
scientific biographies of Bose$^{23}$ and Saha$^{24}$---describing the
scientific achievements of all of them against the backdrop
of physics of their time.  Venkataraman has been much more
critical than a typical biographer of an Indian cultural icon.
An account of Raman's scientific work at Calcutta was 
prepared by S.~N.\ Sen$^{25}$, the doyen of history of science
research in India, who also edited a volume on Saha's life
and works on the occasion of his 60th birthday$^{26}$.  Since Saha
himself went through this material, the account given in this
volume can be taken to be the authorized account of Saha's life
and works as he would have liked to be passed on to posterity. 
Finally, Rajinder Singh has written a study on the discovery
of the Raman effect and its international impact, leading
to the Nobel Prize$^{27}$.

As I have pointed out, our interest is to reconstruct
the history of an extraordinary epoch rather than of individuals.  
One may naively think that we merely take the materials
from the above-listed biographies pertaining to the
epoch of our interest and combine these together to reconstruct the history
of this epoch.  A quick perusal of the biographies listed
above shows that this will not do---one simple reason being
that most of these biographies have only very limited amount
of material concerning the questions of what made this 
extraordinary epoch possible and why it faded away so
quickly. For example, in Venkataraman's biography of Raman
running to more than 500 pages$^{22}$, only about 20 pages are
devoted to events during Raman's golden years when he was
working in Calcutta (1907--33).  In the cases of many famous
persons (including Jesus Christ), it is much easier to get
information about them after they became important public
figures.  It is much harder to gather information about early
stages of their lives before they became famous. I should
mention that there are some valuable studies which cover
the later lives of our protagonists.  Robert Anderson has
published a detailed and thorough study of Saha as an
institute-builder by comparing him with the other 
institute-builders of India of that time (Bhabha, Bhatnagar)$^{28}$.
Just like this study, an interesting study of Raman and
Saha from a feminist viewpoint 
by Abha Sur$^{29}$ also focuses on the later phases of their
lives when they were public figures.

Biographies of famous persons are secondary source materials.
Do we have enough of more primary source materials on which
we can rely in order to reconstruct our history?  To understand
intellectual creativity, often an account of the external
events is not sufficient.  We need to know how the creative person responded
to various influences and what might have been going through
his mind to enable him to be so creative. Unless we have
a record of what a great man thought at a certain time, we
have to guess what might have gone through his mind based on
the available data.  The personal papers of the great 
man---private and official letters, memoirs, recorded 
speeches---often constitute the primary source materials
that allow us a glimpse into the great man's mind and give
us a key to understanding the creative process.  Reminiscences
by contemporaries constitute another valuable primary source
material.  Do we have such source materials for Raman, Bose
and Saha available to us?  Thanks to Saha's children, his
personal papers have been preserved.  Saha himself was an
organized man and kept all his personal papers systematically
arranged.  At one stage a few years after his untimely death, his 
children had to decide what should be done with these personal
papers.  They decided to deposit the personal papers in
the archive of Nehru Memorial Museum and Library, New Delhi,
after making two copies of the entire set of papers---one set
of which is kept at the Saha Institute of Nuclear Physics in 
Calcutta and the other set at the home of Saha's daughter Chitra Roy.
Each set consists of a few thousand pages.
To the best of my knowledge, there has been only one detailed
study of Saha's personal papers (in Bengali) by Atri 
Mukhopadhyay$^{30}$. Not much of the material in the Saha
archive has ever been published.  Unfortunately Bose was
the extreme opposite of Saha and never kept his personal
papers systematically.  Very little of his personal papers seem
to have survived.  However,     
in his old age, Bose was quite fond of reminiscing about
his youth.  He never wrote a long coherent autobiographical
account.  But his reminiscences are scattered through many
pieces---mostly in Bengali.  A few of these pieces were
translated into English on the occasion of Bose's birth
centenary.  But one has to browse through Bose's collected
writings in Bengali$^{12}$ for the other autobiographical clippings.
Perhaps the fate of Raman's personal papers is the most
intriguing.  There is enough evidence that Raman also kept
his personal papers organized very systematically---just
like Saha.  What happened to his personal papers?  Nobody seems
to know an answer to this question!  I have personally enquired
at the five organizations with which Raman had been 
associated---Indian Association for the Cultivation of
Science, Calcutta University, Indian Institute of Science,
Raman Research Institute, Indian Academy of Sciences.  Apart
from some stray documents, none of these organizations has
a substantial systematic collection of Raman papers. Unfortunately,
in India, for a long time there was no awareness that such
materials are of inestimable historical value and should be
preserved properly. I wonder if this is due to our Vedantic
view that life is a transitory illusion! Every historian
knows that records of ancient India are much sparser
compared to the records of other ancient civilizations.  
Only within the last few years, several Indian organizations
have at last started building archives to preserve their
historical records.  In fact, people in all the organizations
with which Raman had been associates are now looking for
Raman materials.  We can only hope that the Raman papers
are still gathering dust in some unknown shelf of a storeroom
in one of these organizations
and will be discovered in the near future.  Right now, we have
to proceed with the assumption that the personal papers of Raman
are largely lost.    

There is one other extremely valuable source for
reconstructing the history of our extraordinary epoch that
has so far been explored very superficially---the records of
the Syndicate and the Senate of Calcutta University.
In the early years of the twentieth century, Calcutta
University used to keep these records in unbelievable
detail.  Very often, minutes of important meetings would
include very detailed statements of the persons who spoke
at these meetings.  For example, the records of the
year 1924 alone run to a whopping 3550 pages!!! It is
certainly not easy to dig out the material you are interested
in from these records.  Recently Calcutta University has
taken the initiative to put these records on the Internet.
In his study
of Raman's scientific work at Calcutta, S.~N.\ Sen$^{25}$ quotes
a few extracts from these reports describing discussions
in which Raman took part.  These extracts quickly dispel
the general perception that Raman was so involved with
his work at the Indian Association for the Cultivation
of Science that he did not take much interest in the affairs
of Calcutta University.  For example, Raman strongly
argued in support of those who wanted to introduce Bengali
at the highest levels of Calcutta University, although Raman
himself could not speak or understand Bengali. I know of
only one scholar who has made extensive studies of the
records of Calcutta University---Dinesh Chandra Sinha,
who retired as the Deputy Registrar of the University.  
A number of articles he wrote on various aspects of Calcutta
University (written in Bengali, though various documents
are quoted in the original English) have been collected 
together in a book$^{31}$.  This book has two articles on Saha
and Raman, quoting several fascinating official letters
between them and the University.  I have already quoted
the letter from Mookerjee to Saha on the financial crisis
of Calcutta University$^5$.  This letter is given in full in
Sinha's book.  Sinha must have obtained these letters from
files kept in some office of Calcutta University.  Most of
Sinha's articles were written more than two decades ago.
Clearly these files existed at that time.  Even after making
several enquiries at Calcutta University, I have not been able to
find any information where these files are kept now or if they
still exist.  When I tried to find contact details of Sinha
in order to meet him personally, I discovered to my regret
that he had passed away just a few months earlier. 
Although Sinha wrote two articles on Saha and Raman reproducing
several official letters to and from them, Sinha unfortunately
did not write a similar article on Bose.  When Santimay
Chatterjee was preparing the biography of Bose on the
occasion of his birth centenary in 1994, he visited Dacca 
University in search of source materials.  He found several 
official letters between Bose and Dacca University neatly
filed. Some of these letters are quoted in the biography
of Bose prepared for his birth centenary$^{19}$.  We do not find
similar official letters between Bose and Calcutta University
quoted in that biography.  My guess is that Chatterjee must
have looked for such letters and could not find them.  So, already
in 1994, Bose's file at Calcutta University was not accessible
to scholars.  Again, with the increased awareness about
the value of such materials, we can only hope that the files
of Raman, Saha and Bose, as well as the files of other famous
professors of Calcutta University of that period, will be
discovered some day from some godown and will thereafter be 
preserved carefully for posterity.

Raman joined the Physics Department of Calcutta University
in 1917 and Saha left for a tour of Europe in September
1919.  By the time he returned, Bose had already left for
Dacca University.  So probably only two or three batches of MSc students
during the years 1917--1919
had the privilege of being taught by Raman, Bose and Saha
simultaneously.  All three of them were young men who still
had not made their famous discoveries and students probably
could not guess that one day these three would be legends of
physics.  What was it like to be an MSc student in the small Physics
Department of Calcutta University at that time? Who taught
what?  Was the teaching exciting? What were the examinations
like?  Can we reconstruct the intellectual atmosphere of
the Department at that time?  From the documents I have so
far looked at, I could not even find out how many students
were there in each batch.  We would like to know the names of
the students in these first batches, whether some of 
them turned out to be professional
physicists and what the others did after MSc. As the Physics
Department of Calcutta University is approaching its centenary,
I find that a few colleagues are interested in finding
answers to these questions. There is a plan of searching
the records of Calcutta University in the next few months
and hopefully we shall have at least partial answers to some
of these questions.  If we could have a reminiscence of the
Physics Department of Calcutta University in those early years,
that would be wonderful.  While I am not aware of anybody who
wrote such a reminiscence, it will probably not be easy to trace 
such a reminiscence even if somebody had written it. I have
already mentioned that Sailen Ghose, who built the MSc physics
laboratory and then could not join the faculty because of his
connection with the revolutionaries, wrote a reminiscence for
the magazine {\em Asia}$^2$.  All my efforts of finding this reminiscence
anywhere in Calcutta failed, although it was reprinted 
in Calcutta in 1992 in the {\em Sailendranath Ghosh Birth
Centenary Commemoration Volume}. Usually such volumes are not
kept and catalogued in libraries. Ultimately I came to 
know that Vivek Bald of Massachusetts Institute of Technology,
who was studying immigrants to America from the Indian
subcontinent, was interested in Sailen Ghose as an eminent
immigrant.  On contacting Bald through e-mail, he told me
that he managed to get those old rare copies of {\em Asia}
through great difficulty and kindly sent me a scanned soft
copy of the Ghose reminiscence. While discussing reminiscences,
I shall mention another unusual document.  Raman's student
K.~S.\ Krishnan, who was involved in the discovery of the
Raman effect and about whom many feel that more credit should
have gone to him, kept a diary describing the work done
in the laboratory.  The mysterious thing about the diary is
that several pages starting from the day on which the Raman effect
was discovered have been torn out.  Nobody knows what happened to
these missing pages. A detailed biography of Krishnan has
been written by D.~C.~V.\ Mallik and S.\ Chatterjee$^{32}$.

It appears that chemists have been more interested in writing
reminiscences than physicists.  The great P.~C.\ Ray---who taught
at Presidency College in Calcutta (where Bose and Saha were his students)
and then joined Calcutta University as the Palit Professor of Chemistry
at the invitation of Asutosh Mookerjee---wrote
a fascinating autobiography both in English and in Bengali$^{33}$.
This invaluable autobiography gives a detailed account of the 
establishment of the Science College of Calcutta University and the subsequent
difficulties it faced from an insider's point of view. Another
interesting reminiscence in Bengali is by P.~C.\ Rakshit$^{34}$, who was
a student of Jnanendra Ghosh (whom Mookerjee mentioned in his letter
to Saha quoted earlier$^5$) in Dacca and has given an intimate portrait of
Bose during his Dacca years.  

In order to put the brilliant epoch of Calcutta physics in proper
historical context, we need to know how modern science grew in India.
A brief account of how Western science started in India has been given
in the seminal work on the history of Indian science by Bose, Sen and
Subbarayapppa$^{35}$. Fuller accounts are given by
Pratik Chakrabarti$^{36}$ and Chittabrata Palit$^{37}$.  J.\ Lourdusamy studied
the four pioneers$^{38}$---Mahendra Lal Sircar, Asutosh Mookerjee, J.~C.\
Bose, P.~C.\ Ray.  Detailed studies of Mookerjee and Ray in Bengali
have been carried out by Shyamal Chakrabarti$^{39-40}$. The authorized biography
of J.~C.\ Bose was by his maverick friend Patrick Geddes$^{41}$.  See also
a recent volume with three long critical essays on Bose$^{42}$. Ashis Nandy
carried out an interesting study of Bose and Ramanujan from a
psychoanalytical point of view$^{43}$.  

Lastly, I should mention the institutional histories. Calcutta
University was established more than half a century before Mookerjee
started the Science College.  On the occasion of the centenary of
Calcutta University in 1957, a group of scholars produced a history
of the University$^{44}$. Before the establishment of the Science College
at Calcutta University,
there were two serious efforts of starting organizations for scientific
research: Indian Association for the Cultivation of Science in Calcutta
(which was discovered by Raman from a tram on his way home)
and Indian Institute of Science in Bangalore (of which Raman became
the first Indian Director in 1933. A history of the first
organization was compiled at the time of its centenary$^{45}$, whereas a history
of the second organization has been written by B.~V.\ Subbarayappa$^{46}$.
Just as the first organization was a dormant sleepy place in its first
few decades, the second organization also had an undistinguished start.
In fact, the failure of Indian Institute of Science to produce any
worthwhile science in its first years was a matter of concern when
Mookerjee was planning the Science College in Calcutta.  The primary
reason for the failure of Indian Institute of Science was that the
British rulers of India were using it as a parking place for second- and third-rate British
academics who could not find any positions in Britain.  To avoid this
from happening in Calcutta University, Mookerjee always insisted that the
donors whose donations created the various named chairs put the condition
that these chairs can only be offered to Indians.  This was another reason
behind the wrath of the British rulers against Calcutta University.

In this survey of source materials, while listing secondary
source materials, I have mainly restricted myself to the most
important book-length studies. There have also been numerous
articles and essays in many journals and magazines that throw
important light on our topic. Since Raman, Bose and Saha are icons
in Indian science, there have also been many short descriptive books
on them---written specially for children and young people.  I have
primarily cited those books which appeared to me to be historically
sound and critical in a scholarly way.

\section{Can we reconstruct the history?}

After this survey of source materials, let us come to the
question whether we can adequately reconstruct the history
of the glorious period of physics research at Calcutta University.
Certain documents which would have been of inestimable value
in reconstructing this history are not available to us---the
most important being the personal papers of Raman and Bose,
as well as the official files of Raman, Bose and Saha with
Calcutta University.  There is a small chance that some of
these documents may be discovered in the coming years. However,
since we are not certain of this, we should make an attempt
at reconstructing the history as best as we can, on the basis
of what we have available to us today. Since we can now
view Raman, Bose and Saha from the distance of a few decades
and most of the persons who had been involved in close scientific
interactions with them are no longer alive, we can now analyze
many events connected with them objectively without hurting the
feelings of any living persons. One thing is clear.  If we
treat Raman, Bose and Saha as faultless supermen that many
of their biographers would like us to believe, then we shall
never be able to address certain important questions---such
as why the brilliant epoch of physics research did not lead
to the establishment of a strong tradition of physics research 
in India. To analyze this issue adequately, we have to treat
them as intellectual giants who nevertheless had many human 
failings.

A few persons who had known Raman, Bose or Saha intimately
are still alive.  One important question is whether we can
obtain valuable data for reconstructing our history by interviewing
them.  I believe that what we can learn from such interviews
is not only of limited value for our study, but can actually be quite
misleading.  These still living persons have all seen these
great men in the twilight years of their lives. We know
that all living beings change with age.  In the cases of
all three of our protagonists, there seem to have been some
critical phase transitions in their personalities---approximately
in the early 1930s in all the cases.  So the three elderly men
whom the still living persons knew were entirely different
persons from the three young men who revolutionized physics.
If we try to interpret the events of 1920s on the basis of
the impression we gather from people who have seen these
great men in the 1950s and 1960s, then we shall be mistaking
apples for oranges.  One simple example will suffice. Although
I have never come across detailed and objective accounts of the 
teaching of Bose and Saha in their younger years, there is
evidence to suggest that they were dedicated and inspiring
teachers in their younger years.  Perhaps the biggest support
in favour of this is Saha's extraordinary textbook on thermal
physics$^{47}$, presumably the best physics textbook ever written
from India.  Having myself written two critically acclaimed
textbooks, I personally know what is involved in writing a
textbook and I cannot believe that a person who was not completely
dedicated in teaching could have written the textbook that
Saha wrote.  However, a few persons who have been taught
by Bose and Saha when they again returned to Calcutta University
are still alive.  Over the years, I had detailed discussions
with at least half a dozen of them (including some who are
not alive today).  I got the uniform opinion that attending
lectures by Bose or Saha was an extremely disappointing
experience.  They both taught in a way as if they had lost all
interest in physics and teaching was an onerous burden in
which their heart was not there.  Bose and Saha, who taught
at Calcutta University in the years immediately preceding
their retirement, were totally different persons from the
young men who took up the teaching of newest developments
in physics as a challenge thrown to them by Asutosh Mookerjee!  

In Indian society, writing anything negative about somebody
who had become a cultural icon is often considered a taboo.
Without exception, all the persons who told me about Bose's
and Saha's unsatisfactory teaching made it clear to me
that they would not like their statements to be recorded or
would not want to be quoted with their names. In scholarly
writing, the standard practice is to provide any information
only with full details about the source of that information.
Although it is still possible to gather anonymous data about
Bose's or Saha's teaching in their later years
from persons who are still alive,
it is not easy to prepare a scholarly account of their
teaching with proper accrediting.  The same difficulty
arises when we try to obtain any negative information about
the personal lives of these great men. Often negative 
developments in personal life affect a person's
intellectual creativity at a deep level and a knowledge
of such developments is often important to explain the
patterns of creativity.  I give an example to demonstrate
how difficult it is to obtain such negative information
about a famous person.  Raman's elder son Raja fell out
with his father, because both of them had dominating 
but incompatible personalities.  During
his adulthood, Raja never had any contacts his father, though
I have heard that his mother secretly used to help him with
money. Except Uma Parameswaran$^{48}$, 
none of the other biographers of Raman even acknowledged
Raja's existence! One gentleman who was close to the Raman 
family told me the following about Raja:
\begin{quote}
He was a very intelligent man and the sharpness of his mind
would be apparent even in casual conversation.  He never
married. He became
a communist in his youth. Since Raman hated communists, that
was the beginning of the breakup of his relation with Raman.
Afterwards Raja converted to Islam.  Although he lived in
the Muslim neighbourhood in a Bangalore locality
called Kodigehalli, 
he and Raman never saw each other
during Raman's final years.  I often used to visit Raja
in his last few years.  He told me that I could visit him
only on one condition: I should never disclose his family
connections to his neighbours. Once when I went to his house,
I found it locked. On enquiring with the neighbours, I came to know
that he had passed away about a week ago. None of the neighbours
knew anything about his family and could not contact any
family members.  He was buried in an unmarked grave in the
nearby Muslim graveyard.
\end{quote}
When I asked this gentleman if I could quote him with his
name in scholarly writing, he told me that he would disown ever
making such statements to me if I quoted him with his name!  

When Bose and Saha returned to Calcutta after their innings
in Dacca and Allahabad respectively, they were legends in
Indian science.  Calcutta has been known for a long time 
as a city where some of the brightest students were interested
in physics.  One would have expected some of these brightest students
to flock around Bose and Saha to learn physics from them and
to work under their guidance.  Something like this does not
seem to have happened.  Purnima Sinha, who did her PhD thesis 
under Bose's supervision and worshipped Bose as a demi-god, 
gave a very bizarre explanation of why the
brightest students did not work with Bose:
\begin{quote}
He never tried to establish a school by gathering the best students
around him.  After all, the best students can always find their own
paths. But who will look after the less capable students? The
great savant therefore took upon himself the job of guiding
such students$^{49}$.
\end{quote}
I am aware of only one frank depiction of the atmosphere of
the Calcutta physics community in the 1940s and 1950s---written
by none other than Amal Raychaudhuri of the Raychaudhuri equation
fame.  It is an extremely frank 30-page autobiographical note
(in Bengali) written when he was convalescing from a serious illness.
Although he lived for three years after this note was written,
it was never published during his lifetime. After his death,
his daughter decided to publish it$^{50}$.  Over the years, I had
heard many of these things written in the autobiographical note 
directly from Professor Raychaudhuri.  He used to tell us that he
wanted promising physics students of the next generation to
know of this history, but he could never think of having this
history published. We are fortunate that we now have this
valuable document. It gives a grim and unflattering portrait
of the Calcutta physics world when Bose and Saha were the living
icons.  We are quite shocked to know of the way Bose and Saha
would interact with young budding physicists of that time. While
Bose never tried to build a research group around himself, Saha
had a thriving research group at Allahabad and several members
of this group became respected physicists afterwards.  We feel
extremely saddened to read the account of Raychaudhuri---especially
when we contrast with it several accounts that we have of Saha's
extraordinary kindness to his students in Allahabad.  How could
an idealistic man who was so inspiring and charismatic in his
younger days change so much?

When several persons of exceptional intellectual ability are
put in the same place, two opposing things can happen.
On the one hand, interactions amongst them can create a vibrant
intellectual atmosphere and enhance the creativity of everybody
around. On the other hand, if resources---laboratory space,
funding, positions---are too limited for several extraordinary
individuals to grow together, that can also lead to a Darwinian
struggle for existence. Both of these things happened within the
Calcutta physics community.  Although Bose and Saha were rivals
in the BSc and MSc classes, they were extremely intimate friends
in their youth and interacted closely.  Apart from writing a
joint paper (Bose's first paper), they together prepared the
first English translation of Einstein's papers on relativity,
barely three years after the publication of Einstein's famous
work on general relativity$^{51}$.
Apparently Bose's famous work on Bose statistics done in Dacca
was also triggered by Saha, who had come to Dacca for some
official work and discussed some recent papers, making Bose
interested in the problem of energy distribution of photons$^{52}$.
Although Bose and Saha never fell out with each other, the
intensity of friendship diminished considerably in their later
lives.  When I asked Saha's daughter Chitra Roy whether Bose
sometimes visited Saha at his home or Saha visited Bose at his
home when they were both in Calcutta in their later years, Chitra
Roy could not recollect any such visits. 

To collect reliable information about the conflicts amongst scientists
is much more difficult than to collect information about their
cooperation.  Both Raman and Saha had strong dominating
personalities.  It is clear that each man admired the other for
the other's physics contributions.  However, there was a clash of egos
between these two titans in the early 1930s that had a disastrous
consequence for Indian science.  I have mentioned about the drastic
changes in the personalities of Raman and Saha.  These changes more
or less coincide with this clash, making us wonder whether this
unfortunate clash was a major contributor in the changes of their
personalities.  It is not easy to find reliable information of exactly
what happened.  Some biographers have squarely taken the side of
their biographees.  For example, Uma Parameswaran$^{20}$, the biographer of
Raman, would like us to believe that Raman was a faultless 
person of childlike simplicity who was not to be blamed for any
conflicts he was involved in.  On the other hand, Atri Mukhopadhyay$^{30}$,
who studied Saha's private papers, projects Saha as an innocent victim
of the conspiracies hatched by others against him.  I find G.\ Venkataraman to be
the {\em only} author who has been impartial and made the following
perceptive comment:
\begin{quote}
Clashes between strong personalities are not uncommon and they are
to be found everywhere and at all times.  But in the academic world,
one seldom witnesses events such as those which occurred in Calcutta.
The basic problem there was that there was not enough room for both the
giants.  If only the country had been rich enough to afford several
faculty and student positions, this conflict might never have gone beyond
minor skirmishes$^{53}$.
\end{quote}

But even Venkataraman reconstructed events ``based on conversations
with persons who had first-hand knowledge'' and not on the basis of 
proper documentation. According to Venkataraman, Saha was interested in
a professorship created at the Indian Association for the Cultivation of
Science and requested the support of Raman, who was the Secretary of the
Association. Raman, who wanted this position to be offered to his
former student Krishnan, wrote back saying that ``lately Saha had not
been much concerned about research; what the Association needed was an
active young man on his way up rather than one who had reached a plateau''$^{54}$.
These letters between Saha and Raman have not been found! Atri Mukhopadhyay
dismissed this version as a story fabricated by Raman's supporters. 
Perhaps the account of the events in a fateful meeting 
at the Indian Association for the Cultivation
of Science that Mallik and Chatterjee$^{55}$ have given is as much as we can 
now establish on the basis of documentary evidence.  But what Saha
wrote in a letter to P.~C.\ Ray indicates that there were more things
happening and the version of Venkataraman may have some truth. Saha's
language in the letter is quite shocking:
\begin{quote}
Prof.\ Raman will have to know that if he does not leave the Science
Association to Bengal, he will have to meet with determined opposition
from me.  Ten years administrative experience at Allahabad have not been
in vain.  He will find a Tartar in me, and you may drop him a hint that 
is not my only trick. I have got many other obstructive tactics up my
sleeve.  This time I placed my cards on the table, but next time, I shall
not give him even this chance.  I will confront him with the difficulties
on the spot and make him dance$^{56}$.
\end{quote}

While these conflicts rose to a crescendo in the early
1930s, there were tussles for scarce resources as soon as
the Science College of Calcutta University was established.
J.~C.\ Bose and Raman were the two outstanding experimental
physicists of that era.  It is unfortunate that they fell out
with each other very soon.  S.~N.\ Bose, who had seen this encounter
from a close distance, reminisced many years later:
\begin{quote}
when we were students, the professors used 
to hold on to their mechanics as some of the most precious
possessions \ldots there was a mechanic with Sir J~C Bose and he
did very fine work---indeed to his amazement. Because Sir C~V Raman
had cast his jealous eyes on this mechanic \ldots he grew furious \ldots
this Sir J~C Bose wouldn't forget and wouldn't forgive$^{57}$.
\end{quote}
In fact,  J.~C.\ Bose was so furious that
he wrote an angry letter to the Vice-Chancellor of Calcutta
University, complaining that Raman was trying to lure away
this mechanic by offering a much higher salary$^{58}$. 
While this might have been acceptable in the American academic
world, this was certainly unacceptable behaviour in India
of that period. Given the fact that J.~C.\ Bose was not employed
at Calcutta University, his writing a letter to the 
Vice-Chancellor clearly shows how disgusted he was.  Several
years later, when Raman was felicitated by the Mayor of Calcutta
on receiving the Nobel Prize, all the important persons in the
Calcutta scientific world were there, except J.~C.\ Bose.
While there could be other simple explanations for this absence
(illness, not being in Calcutta on that day), it appears that
the relationship between J.~C.\ Bose and Raman always remained
strained.  However, when Sommerfeld visited India, J.~C.\ Bose
accompanied him for a visit to Raman's lab$^{59}$.  Clearly J.~C.\
Bose did not want a distinguished scientist from abroad to know
of their differences.

Abha Sur has written about a seminar at Harvard in 1998:
\begin{quote}
[The] speaker showed a group photograph of scientists and
went about identifying all the European and American scientists
in the picture. In cases where the speaker did not know
the identity of the scientist, the audience was asked for
help. There was C.~V.\ Raman right in the middle of the
front row of the picture in his big turban, conspicuous
in his difference from the rest of the scientists, but the
speaker's pointer simply slid over his imposing personality
without a pause, without any hesitation whatsoever.  The otherwise
alert and inquisitive audience seemed not to mind the omission$^{60}$.
\end{quote}
Although every physicist around the world is expected to know of
the famous works of physics connected with the names of Raman,
Bose and Saha, the extraordinary story of how they created
their physics fighting against all odds remains largely unknown
and untold to the international physics community. 
This story needs to be told,
even if we do not have all the source materials to reconstruct the
full history. While this is an uplifting 
and dramatic story of how the human spirit
can conquer many obstacles, this is also a story of
great sadness: a story of how the human failings of the extraordinary
individuals who created this glorious epoch of physics---all
fiercely patriotic and idealistic in their youth---coupled
with forces in the colonial setup that was beyond their control,
finally extinguished the light that shone with such brilliance for
a short epoch.  

\section*{Notes}

1. This meeting has been described by S.~N.\ Bose in an autobiographical
article written in Bengali: {\em Satyendra Nath Basu Rachana Sankalan}
(Bangiya Vijnan Parishad, Calcutta, 1980), pp.\ 226--227.  Asutosh Mookerjee's tenure
as Vice-Chancellor ended in 1914, though he was again appointed Vice-Chancellor
a few years later.  Most probably this meeting took place in 1915 when
Ghose, Bose and Saha completed their MSc. Then Mookerjee was not the Vice-Chancellor
exactly at that time, although he was still very much giving shape to the new activities
initiated at Calcutta University. 

2. Sailendra Nath Ghose, ``An Indian Revolutionary'', {\em Asia}, Part I: p.\ 586 (July 1927);
Part II: p.\ 671 (August 1927); Part III: p.\ 737 (September 1927).

3. S.\ Ramaseshan and C.\ Ramachandra Rao, {\em C.~V.\ Raman: A Pictorial Biography}
(Indian Academy of Sciences, Bangalore, 1988), quote at pp.\ 15--16.

4. {\em Sir Asutosh Mookerjee: A Tribute} (University of Calcutta, 2013), quote at p.\ 198.

5. Dinesh Chandra Sinha, {\em Prasanga: Kolkata Biswabidyalay} (University of Calcutta, 
2007), quote at p.\ 250. 

6. A fascinating account of Ramanujan's life has been given by Robert Kanigel, {\em The
Man Who Knew Infinity: A Life of the Genius Ramanujan} (Washington Square Press, 1991).
This book portrays the academic environment in another region of India at the beginning
of the twentieth century.

7. Amal Kumar Raychaudhuri, ``A century of progress'' (unpublished manuscript). 
Courtesy: Subinay Dasgupta.

8. S.\ Ramaseshan (editor), {\em Scientific Papers of C V Raman}, in 6 volumes 
(Indian Academy of Sciences, Bangalore, 1988).

9. {\em S N Bose: The Man and His Work. Part I: Collected Scientific Papers} (S N Bose National Centre for
Basic Sciences, Calcutta, 1994).

10. {\em Collected Scientific Papers of Meghnad Saha} (Council of Scientific and
Industrial Research, Government of India, New Delhi, 1969).

11. Santimay Chatterjee (editor), {\em Collected Works of Meghnad Saha}, in 4 volumes 
(Orient Longman, 1982--1993)

12. {\em Satyendra Nath Basu Rachana Sankalan}
(Bangiya Vijnan Parishad, Calcutta, 1980).

13. Santimay Chatterjee (editor), {\em Meghnad Rachana Sankalan} (Orient Longman, 1986).

14. D.~S.\ Kothari, ``Meghnad Saha 1893--1956'', {\em Biographical Memoirs 
of Fellows of the Royal Society} {\bf 5}, 217 (1959).

15. S.\ Bhagavantam, ``Chandrasekhara Venkata Raman 1888--1970'', {\em Biographical Memoirs 
of Fellows of the Royal Society} {\bf 17}, 564 (1971).

16. Jagdish Mehra, ``Satyendra Nath Bose 1 January 1894-–4 February 1974''. {\em Biographical Memoirs 
of Fellows of the Royal Society} {\bf 21}, 116 (1975).

17. Santimay Chatterjee and Enakshi Chatterjee, {\em Satyendra Nath Bose} (National Book Trust, New Delhi, 1978).

18. Santimay Chatterjee and Enakshi Chatterjee, {\em Meghnad Saha} (National Book Trust, New Delhi, 1984).

19. ``S N Bose: Life'', in {\em S N Bose: The Man and His Work. Part II: Life, Lectures and Addresses,
Miscellaneous Pieces} (S N Bose National Centre for Basic Sciences, Calcutta, 1994).

20. Uma Parameswaran, {\em C.\ V.\ Raman: A Biography} (Penguin Books India, 2011).

21. Several of S.\ Ramaseshan's articles on Raman, though not all of them, are collected in
S.\ Ramaseshan and C.\ Ramachandra Rao, {\em C.~V.\ Raman: A Pictorial Biography}
(Indian Academy of Sciences, Bangalore, 1988).

22. G.\ Venkataraman, {\em Journey into Light: Life and Science of C.~V.\ Raman} (Indian Academy
of Sciences, Bangalore, 1988).

23. G.\ Venkataraman, {\em Bose and His Statistics} (Universities Press, Hyderabad, 1992).

24. G.\ Venkataraman, {\em Saha and His Formula} (Universities Press, Hyderabad, 1995).

25. S.\ N.\ Sen, {\em Prof.\ C.\ V.\ Raman: Scientific Work at Calcutta} (Indian Association
for the Cultivation of Science, Calcutta, 1988).

26. S.\ N.\ Sen (editor), {\em Professor Meghnad Saha: His Life, Work and Philosophy} (Meghnad
Saha Sixtieth Birthday Committee, Calcutta, 1954).

27. Rajinder Singh, {\em Nobel Laureate C.\ V.\ Raman's Work on Light Scattering: Historical
Contributions to a Scientific Biography} (Logos Verlag, Berlin, 2004).

28. Robert S.\ Anderson, {\em Nucleus and Nation: Scientists, International Networks, and Power
in India} (The University of Chicago Press, Chicago, 2011).

29. Abha Sur, {\em Dispersed Radiance: Caste, Gender, and Modern Science in India} (Navayana, New Delhi, 2011).

30. Atri Mukhopadhyay, {\em Abinash Meghnad: Bijnan Samaj Rashtra} (Anushtup, Calcutta, 2012).

31. Dinesh Chandra Sinha, {\em Prasanga: Kolkata Biswabidyalay} (University of Calcutta, 
2007).

32. D.\ C.\ V.\ Mallik and S.\ Chatterjee, {\em Kariamanikkam Srinivasa Krishnan: His Life and
Work} (Universities Press, Hyderabad, 2012).

33. P.\ C.\ Ray, {\em Life and Experiences of a Bengali Chemist}, 2 volumes (Chuckervertty,
Chatterjee and Co., Calcutta, 1932 and 1935).

34. Pratulchandra Rakshit, {\em Periye Elem} (Sharat Book Distributors, Calcutta, 1992).

35. D.\ M.\ Bose, S.\ N.\ Sen and B.\ V.\ Subbarayappa, {\em A Concise History of Science
in India} (Indian National Science Academy, New Delhi, 1971).

36. Pratik Chakrabarti, {\em Western Science in Modern India: Metropolitan Methods, Colonial
Practices} (Permanent Black, Delhi, 2004).

37. Chittabrata Palit, {\em Scientific Bengal: Science, Technology, Medicine and Environment
under the Raj} (Kalpaz Publications, Delhi, 2006)

38. J. Lourdusamy, {\em Science and National Consciousness in Bengal 1870--1930} (Orient 
Longman, New Delhi, 2004).

39. Shyamal Chakrabarti, {\em Oitihya Uttaradhikar o Bijnani Prafullachandra} (Sahitya Sansad,
Calcutta, 2009).

40. Shyamal Chakrabarti, {\em Sikshar Shirsho Sthapati: Sir Asutosh Mukhopadhyay} (Jnan
Bichitra, Agartala, 2014).

41. Patrick Geddes, {\em An Indian Pioneer of Science: The Life and Work of Sir Jagadis C.\
Bose} (Longmans, Green and Co., London, 1920).

42. D.\ P.\ Sengupta, M.\ H.\ Engineer and V.\ A.\ Shepherd, {\em Remembering Sir J C Bose}
(World Scientific, Singapore, 2009).

43. Ashis Nandy, {\em Alternative Sciences} (Allied Publishers, New Delhi, 1980).

44. {\em Hundred Years of the University of Calcutta} (University of Calcutta, 1957).

45. {\em A Century} (Indian Association for the Cultivation of Science, Calcutta, 1976).

46. B.\ V.\ Subbarayappa, {\em In Pursuit of Excellence: A History of the Indian Institute
of Science} (Tata McGraw-Hill, New Delhi, 1992).

47. M.\ N.\ Saha and B.\ N.\ Srivastava, {\em A Treatise on Heat} (The Indian Press, 
Allahabad, 1931).

48. Parameswaran, {\em C.\ V.\ Raman}, pp.\ 188--190.

49. Purnima Sinha, {\em Bijnansadhanar Dharay Satyendranath Basu} (Visva-Bharati, Calcutta, 1981), 
quote at p.\ 70 (my translation from Bengali).

50. Amal Kumar Raychaudhuri, {\em Atmajijnasa o Anyanya Rachana} (Sharat Book Distributors, Calcutta, 2007).

51. {\em The Principle of Relativity: Original Papers by A.\ Einstein and H.\ Minkowski}, Translated
into English by M.~N.\ Saha and S.~N.\ Bose (Calcutta University, 1920).

52. P.\ K.\ Roy, ``Thermal Radiation Laws, Bose Statistics and its Immediate Impact'', {\em Science
and Culture} {\bf 40}, p.\ 293 (January 1974).

53. Venkataraman, {\em Journey into Light}, quote at p. 59.

54. Venkataraman, {\em Journey into Light}, quote at p. 58.

55. Mallik and Chatterjee, {\em Kariamanikkam Srinivasa Krishnan}, pp.\ 149--151. See also
Mukhopadhyay, {\em Abinash Meghnad}, pp.\ 116--126, for a somewhat biased account.

56. Chakrabarti, {\em Oitihya Uttaradhikar o Bijnani Prafullachandra}, quote at p.\ 341.

57. {\em S N Bose: The Man and His Work. Part II}, quote at p.\ 194.  

58. Singh, {\em Nobel Laureate C.\ V.\ Raman's Work on Light Scattering}, p.\ 17.

59. Venkataraman, {\em Journey into Light}, p.\ 55.

60. Sur, {\em Dispersed Radiance}, quote at p.\ 17.

\end{document}